\documentclass[floatfix,twocolumn,
aps,prl,showpacs,showkeys,amsmath,amssymb,%
]{revtex4} 

\usepackage{dcolumn}
\usepackage{bm}
\usepackage{graphicx}

\def\half{\frac{1}{2}}
\def\slash#1{\, /\kern-0.6em{#1}}

\begin{document}


\preprint{{hep-th/0312112}}

\title{Parallel transport on non-Abelian flux tubes}
\author{Amitabha Lahiri}
\email{amitabha@boson.bose.res.in}
\affiliation{S. N. Bose National Centre for Basic Sciences, \\
Block JD, Sector III, Salt Lake, Calcutta 700 098, INDIA\\ 
amitabha@boson.bose.res.in}
\date{\today}

\begin{abstract}
I propose a way of unambiguously parallel transporting fields on
non-Abelian flux tubes, or strings, by means of two gauge fields.
One gauge field transports along the tube, while the other
transports normal to the tube. Ambiguity is removed by imposing an
integrability condition on the pair of fields.	The construction
leads to a gauge theory of mathematical objects known as Lie
2-groups, which are known to result also from the parallel
transport of the flux tubes themselves. The integrability condition
is also shown to be equivalent to the assumption that parallel
transport along nearby string configurations are equal up to
arbitrary gauge transformations. Attempts to implement this
condition in a field theory leads to effective actions for two-form
fields.

\end{abstract}


\maketitle

Just as a gauge field couples to the world line of a charged
particle, a two-form field couples to the world sheet of a string.
But unlike particles with non-Abelian charge, which couple to gauge
fields, non-Abelian strings or flux tubes pose a mathematical
hurdle. The crux of the problem is to consistently define a notion
of surface holonomy, i.e., to assign elements of a group to
surfaces swept out by parallel transport of these strings. Simply
put, the holonomy of a surface is trivial unless the holonomy group
is Abelian~\cite{Teitelboim:1986ya}, essentially because there is
no canonical notion of surface ordering. Recently there have been
fresh attempts to overcome this problem by circumventing it. The
proposal is that the proper geometrical setting for non-Abelian
string dynamics is a category theoretic generalization of principal
fiber bundles called gerbes~\cite{Breen:2001ie}. Further, while the
symmetry transformations of fields which describe particles form
groups, for non-Abelian strings a more generalized notion of
symmetry is needed, encoded in structures known as Lie 2-groups or
Lie crossed modules. Different authors have given descriptions of
these structures~\cite{Attal:2001,Attal:2002ix,Chepelev:2001mg,Baez:2002jn,Hofman:2002ey,Pfeiffer:2003je,Girelli:2003ev}.

In these discussions, categorical groups and gerbes have been
associated with two-form gauge fields. For example, in the 2-group
construction of surface holonomy, a one-form field and a two-form
field, valued in the two Lie algebras, are introduced for parallel
transport of a charged string. The two-form couples to the face,
while the one-form couples to the edges, for a holonomy valued in
the Lie 2-group.  This construction leads to a solution of surface
holonomy problem~\cite{Baez:2002jn,Pfeiffer:2003je}.  Non-Abelian
two-form fields also appear in string theory when two or more
D-branes coincide. The geometrical setting for these fields is
provided by gerbes, which are like principal fiber bundles of
categorical groups~\cite{Baez:2002jn}.	Another relevant example is
that of discrete torsion. The action of the orbifold group on the
two-form field in string theory gives rise to discrete torsion. The
contribution of this to the string partition function comes from
the holonomy of the two-form~\cite{Sharpe:2000ki}.

In this note I point out that Lie 2-groups, which at first sight
may seem to be rather esoteric mathematical objects, appear very
naturally in any theory with charged one-dimensional objects like
flux tubes or strings. It is not necessary to start with a two-form
field and construct parallel transport of strings, i.e. surface
holonomy, as has been done so far. My approach will be to consider
instead parallel transport of {\em fields}, both along and between,
nearby flux tube configurations. I will show that the usual notion
of holonomy along paths, applied to paths along and between flux
tubes, give rise to Lie 2-groups. Since fields and their parallel
transport by gauge fields are ubiquitous in physics, this
construction may be useful in many different contexts where flux
tubes and strings appear in field theory, not only in string
theory, and not only if a two-form field is present a priori.

Even though I do not consider parallel transporting a string
itself, a two-form field will appear naturally as a Lagrange
multiplier in this construction. A simple argument, typical in
quantum theory, then produces an action for this field.
I will also show that the composition of parallel transport of {\em
fields} living on strings follows the rules of a Lie 2-group, or
more precisely those of a Lie crossed module, in exactly the same
manner as the composition of parallel transport of the {\em
strings} themselves. In what follows, I will use the terms Lie
2-group and Lie crossed module interchangeably.

What is a Lie crossed module? In a simplified definition, it
consists of two Lie groups $G$ and $H$, with a homomorphism $t:
H\to G$ and an action of $G$ on $H\,,$ which may be written as
$ghg^{-1}\,,$ where $g\in G$ and $h\in H\,.$ The homomorphism is
required to preserve this action, i.e.
\begin{eqnarray}
t(ghg^{-1}) = g\,t(h)\,g^{-1}\,,\qquad t(h)\,h'\,t(h)^{-1} = h h'
h^{-1}\,.
\label{1202.lcmdef}
\end{eqnarray}
Roughly speaking, the idea is to associate elements of $G$ and $H$
to string configurations such that an element $g\in G$ is `along'
the string configuration in some sense, while an element $h\in H$
`compares' two configurations between the same endpoints. I will
loosely call this a string `carrying' $(h, g)\,,$ keeping in mind
that actually two configurations are needed to define $(h, g)\,.$
Then composition of paths along the string (horizontal composition)
corresponds to multiplication in $G$, while composition of paths in
the space of string configurations with fixed endpoints (vertical
composition) corresponds to group composition in
$H$~\cite{Baez:2002jn,Pfeiffer:2003je,Girelli:2003ev}.

One can define a corresponding Lie 2-group such that horizontal
composition of a directed string, carrying $(h_1, g_1)$ to another
one carrying $(h_2, g_2)$ is the group composition in the
semi-direct product $H\rtimes G\,,$
\begin{eqnarray}
(h_1, g_1)\cdot (h_2, g_2) = (h_1 g_1 h_2 g_1^{-1}, g_1 g_2)\,.
\label{1202.hcomp}
\end{eqnarray}
Vertical composition of one configuration carrying $(h, g)$ with
another between the same endpoints and carrying $(h', g')$, which
is defined if and only if $g' = t(h)g\,,$ is given by
\begin{eqnarray}
(h, g)\circ (h', g') = (h' h, g)\,.
\label{1202.vcomp}
\end{eqnarray}

The transition from geometry to physics is natural when associating
paths to group elements. This is done by taking a gauge field, or
connection, valued in the corresponding Lie algebra, and
calculating its (untraced) holonomy, or `path-ordered exponential',
along the path.	 In order to use this procedure for strings, I will
think of horizontal composition as composition of holonomies along
the string, and vertical composition as composition for holonomies
between two string configurations.

For composition of strings carrying two groups, I could then think
of two gauge fields attached to the string. One of the gauge fields
parallel transports along the string, while the other parallel
transports between string configurations. This is different from
the picture involving Lie crossed modules, in which the recent
attempts to define parallel transport associated an element of $H$
with the surface between two string configurations, by introducing
a two-form gauge field which transports along surfaces. Then
constructing a gauge theory for Lie crossed modules is the same as
giving a definition of surface holonomy, because composition of
elements of Lie 2-groups can be made equivalent to composition of
holonomies along both surfaces and edges.  It will be seen that the
holonomies of the two gauge fields produce the same structure of a
Lie 2-group. For the moment, I will ignore the 2-group construction
of surface holonomy, and derive results in terms of gauge fields,
coming back to the similarity at the end.

I will restrict my considerations to infinitesimal parallel
transports, i.e.\, holonomies along infinitesimal paths. Parallel
transports are done along the edges of an infinitesimal square, in
which the top and bottom edges belong to infinitesimally close
configurations of the string, while the side edges are
infinitesimal paths connecting the configurations. Corresponding
group elements are of the generic form $\exp(\epsilon A) \sim 1 +
\epsilon A\,,$ where $\epsilon$ is the length (vector) of an edge.
Parallel transport from one corner of the square to another corner
should be unambiguous.

\begin{figure}[htbp]
\includegraphics[width=.36\columnwidth]{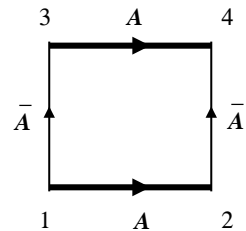}%
\caption{Reference loop for doing parallel
transports.\label{1202.f.square}}
\end{figure}

If there were only one gauge field present, all parallel transports
would be done with the help of this field $A\,.$ Then parallel
transporting a field around an infinitesimal square produces a term
proportional to the curvature $F = {\rm d}A + A\wedge A\,.$ If
${\cal P}_{ab}$ denotes the operator for parallel transport from
$a$ to $b\,$ in Fig.~\ref{1202.f.square}, ${\cal P}_{24}{\cal
P}_{12} = {\cal P}_{34}{\cal P}_{13}$ implies that $F = 0\,.$ This
is the well-known integrability condition --- removing the
ambiguity in parallel transport forces the connection $A$ to be
flat. An equivalent way of stating this is that if it does not
matter where the string is, i.e., if parallel transporting between
two points along one path gives the same result as parallel
transporting along an infinitesimally near path, $A$ must be
flat. Then the corresponding gauge theory is one of flat
connections only. However, motivated by the earlier discussion that
non-Abelian symmetries of strings has the structure of Lie
2-groups, I can relax my assumptions slightly.

Given a smooth string, it is possible to distinguish edges which
lie along it from edges which are normal to it.	 Let parallel
transports along the string be done with the help of the gauge
field $A\,,$ and let parallel transports normal to the string be
done with a different gauge field $\bar A\,.$ To begin with,
suppose the two Lie algebras are identical.
Then I can write $\bar A = A + V\,,$ where $V$ is some one-form
belonging to the same Lie algebra as $\bar A\,.$ The effect of
parallel transporting around an infinitesimal square is then no
longer only the curvature. A simple calculation shows that removing
the ambiguity in parallel transporting around an infinitesimal
square requires
\begin{eqnarray}
F + \half\, {\rm d}_A V \equiv F + \half\, ({\rm d}V + A\wedge V +
V\wedge A) = 0\,.
\label{1202.weak}
\end{eqnarray}

Another way of thinking about this result is to consider, as
before, parallel transport along the bottom edge of the square. A
weaker version of the integrability condition is that going around
the other three edges gives the same result, but only up to gauge
transformations. In the notation used earlier, this condition can
be written as $({\cal P}_{24})^{-1}{\cal P}_{34}{\cal P}_{13} =
a^{-1}{\cal P}_{12}a\,,$ for some group element $a\,,$ where now
all the parallel transports are done with the same connection
$A\,.$ Since all the paths are infinitesimal, and since $a$ should
be the identity when the side edges are collapsed so that the top
and bottom edges coincide, I can write $a \sim 1 + \epsilon\cdot
V\,.$ This leads to Eq.~(\ref{1202.weak})\,, which determines $A$
in terms of $V$ in principle. In this picture, $V$ is not a
conenction, but an arbitrary one-form field valued in the Lie
algebra.

Note that even though the two connections $A$ and $\bar A$ (or the
fields $A$ and $V$ in the alternative picture above) turned out to
be related, the reason for working with two connections has not
disappeared. If only one connection $A$ were used to describe the
parallel transport, I would get $F = 0$ everywhere, making parallel
transport a trivial operation. However, there is a priori no reason
to use only one connection, as there are two inequivalent
directions on the `soap bubble' between two configurations of the
flux tube. So there is no reason why parallel transport must be
trivial, since I can now use two different connections in the two
directions. Then the parallel transport is encoded in the pair
$(h,g)$ which comes from path-ordered exponentials of these two
connections along the two directions.  The collection of these
pairs form a Lie 2-group, as will be discussed in detail later. The
`alternative picture' above has a slightly different justification,
that parallel transport along two configurations should be gauge
equivalent rather than equal. This leads to the same expression for
parallel transport.

Note also that $a$ need not be in the same group as the holonomies
of $A\,,$ but can live in a subgroup of the latter. If this is a
normal subgroup, generators of this group commute with the
remaining generators of the Lie algebra. Then Eq.~(\ref{1202.weak})
is satisfied only in this subalgebra, while the connection is flat
along the other generators. In this case, and even in general, I
ought to consider isomorphisms of one algebra with an invariant
subalgebra of the other, rather than blithely equating objects in
different spaces. That is, I really ought to write $dt(V)$ instead
of $V$ in Eq.~(\ref{1202.weak}) and $(t(h), g)$ instead of $(h, g)$
for the parallel transport. This notation will allow comparison
with the literature easier, but I will continue to keep the
isomorphism implicit to avoid cluttering equations.

The condition in Eq.~(\ref{1202.weak}) is a constraint on the
fields in the background on which the flux tube propagates. Any
effective field theory of these fields will have to incorporate the
constraint. So it helps to think of the constraint as a statement
of gauge equivalence of parallel transports along nearby
configurations of a flux tube. For, now the integrability condition
can be stated as follows.  Parallel transports along two nearby
string configurations, which share the same endpoints, are equal up
to arbitrary gauge transformations. The adjoint vector field $V\,,$
which exponentiates to the group element $a\,,$ may have any
configuration allowed by Eq.~(\ref{1202.weak})\,. Since there is no
reason a priori to prefer any one configuration, I should sum over
all possible configurations of $V\,.$

Different field configurations can be summed over in a path
integral of the form
\begin{eqnarray}
Z = \int {\cal D}A\, {\cal D}V\, \delta\Big[F + \half\,{\rm d}_A
V\Big]\,e^{iI}\,.
\end{eqnarray}
Here the $\delta$-functional enforces the constraint, and $I$ is
some action which can involve both $V$ and the gauge field $A\,.$
To begin with, let me rewrite the $\delta$-functional using a
Lagrange multiplier field $B\,,$ to get
\begin{eqnarray}
Z = \int {\cal D}A\, {\cal D}V\, {\cal D}B\,  e^{{ iI
+ i\int B\wedge( F + \half\,{\rm d}_A V )}}\,,
\end{eqnarray}
The field $B$ is a two-form in four dimensions, and integrating
over $V$ will lead to an effective action for the non-Abelian
two-form $B$. Geometrical considerations, as discussed so far, do
not lead to a specific form of the action $I$. Different choices
for the action $I$ will obviously lead to different field theories
involving $B$.

For example, suppose I choose the simplest possible action, namely
that of pure gauge theory. Then the path integral takes the form
\begin{eqnarray}
Z = \int {\cal D}A\, {\cal D}V\, {\cal D}B\,  e^{{ -i\int
\half F^2  - B\wedge( F + \half\,{\rm d}_A V )}}\,,
\end{eqnarray}
where $F^2 = F\wedge*F\,,$ and gauge indices have been summed over
and hidden. $V$ can be integrated out from this, producing a
constraint ${\rm d}_A B = 0\,.$ The Lagrangian of the resulting
effective theory is ${\cal L}_{\rm eff} = -\half F\wedge*F +
B\wedge F\,.$ This is the usual Lagrangian for topological field
theory in four dimensions, as $F$ can be set to zero using the
equation of motion for $B\,,$ and the constraint ${\rm d}_A B =
0\,$ is then the equation of motion for $A\,.$ The same effective
action results from setting $V = 0\,.$

Another interesting effective action appears if I note that instead
of summing over all configurations of $V$ with equal weight, I
could choose to take a Gaussian average. That is, I could assume
that $V$ is peaked around zero, so that the connection for parallel
transporting a field away from the string is peaked around the
value of the connection for parallel transporting along the
string. The averaging is done by adding a term proportional to
$V^2$ to the exponent in the path integral.
Then the partition function is
\begin{eqnarray}
Z = \int  {\cal D}A\, {\cal D}V\, {\cal D}B\,  e^{{ -i\int
\half F^2 + \half m^2 V^2 - B\wedge( F + \half\,{\rm d}_A V )}}\,,
\end{eqnarray}
where I have introduced a constant $m$ of mass dimension one so
that all terms have the same dimension. Integrating over $V$ shows
that this is the partition function for the action
\begin{eqnarray}
I_{\rm eff} = \int -\half F\wedge *F - \half H\wedge *H + m B\wedge
F\,,
\label{1202.effac}
\end{eqnarray}
where $B$ has been redefined as $mB\,,$ and $H = {\rm d}_A B\,.$

These actions should not come as a surprise. Some theory of
two-form fields is expected in the effective field theory of string
backgrounds from general considerations. For example, the effective
action for `global' vortex strings is that of a free two-form
field, while cosmic strings attached to magnetic monopoles give
rise to an effective action of the form
Eq.~(\ref{1202.effac})~\cite{Vilenkin:1986ku,Lee:1993ty}.
Similar actions also arise from generalizing the notion of a fiber
bundle, which describes the geometry of interactions of point
particles, to a category expected to play a similar role for
strings.  It is possible to introduce the pair $(A, B')$ as the
generalized connection on a `principal
2-bundle'~\cite{Baez:2002jn}, where $A$ is valued in the algebra
${\cal G}$ and $B'$ is valued in the algebra ${\cal H}\,.$ The
one-form $A$ provides holonomies along edges, while the two-form
$B'$ provides holonomies for surfaces. Then one can write actions
similar to the ones above. Here I have shown that at least in four
dimensions, even if transport of strings is not included in the
original construction, an effective action for a non-Abelian
two-form appears naturally.  I have used a prime to distinguish
between the $B$-field in that picture from the one described
above. It is important to note that the $B$-fields in the two
pictures do not seem to be related by a local transformation.

\begin{figure}[htb]
\includegraphics[bb=40 0 440 210, width=.8\columnwidth]{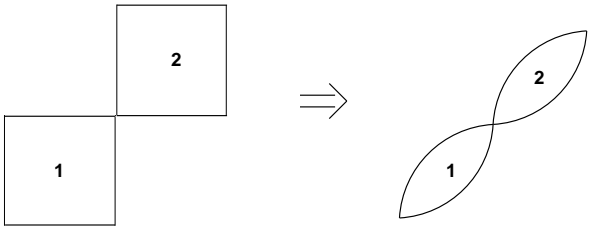}%
\caption{Diagonal composition as horizontal
composition.\label{1202.f.bigon}}
\end{figure}

So far in my construction, the references to Lie 2-groups, Lie
crossed modules and surface holonomy may have looked somewhat
arbitrary or far-fetched. It is time to establish a closer
relationship.  First note that parallel transport around an
infinitesimal square is defined by two objects, one for the
horizontal transport and one for the vertical transport.
Eq.~(\ref{1202.weak}) ensures that parallel transport of a field
from any corner to any other is unambiguous. When two squares are
joined, any paths may be taken from one corner to another, but some
paths make it easier to compare the construction given here with
the Lie 2-group appearing in surface holonomy.

\begin{figure}[thb]
\includegraphics[bb=0 0 400 210, width=.8\columnwidth]{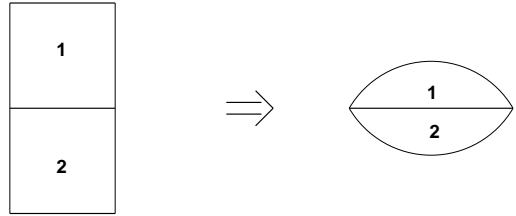}%
\caption{Vertical composition.\label{1202.f.vert}}
\end{figure}

Let me associate a group element $g\in G$ to dragging left along
the top edge and $h \in H$ to dragging down along the left edge.
Then parallel transport from the top right corner to the bottom
left corner of a square with $(h, g)$ requires the combination
$hg\,,$ which is well-defined if the two Lie algebras are
isomorphic, or if one is a subalgebra of the other. If two squares
are joined at a corner as in Fig.~\ref{1202.f.bigon}, parallel
transport from the top right corner to the bottom left corner is
done by $h_1 g_1 h_2 g_2\,,$ which is the same as that for a square
with $(h_1 g_1 h_2 g_1^{-1}, g_1 g_2)\,.$ This operation of joining
at a corner corresponds to horizontal composition of bi-gons as
described in Refs.~\cite{Baez:2002jn,Pfeiffer:2003je}\,. If two
squares are joined along an edge, say a horizontal edge as in
Fig.~\ref{1202.f.vert}, dragging left and down requires a parallel
transport of $h_2 h_1 g_1\,,$ same as a rectangle with $(h_2 h_1,
g_1)\,.$ This operation corresponds to vertical composition of
bi-gons. Comparison with Eq.~(\ref{1202.hcomp}) and
(\ref{1202.vcomp}) shows that the construction in terms of two
connections is indeed a `gauge theory' of Lie 2-groups. In fact,
these pictures and corresponding composition rules are the same as
those given in
Refs.~\cite{Baez:2002jn,Pfeiffer:2003je,Girelli:2003ev},
although the gauge theories obtained here are not exactly the same
as those found there, where the two-form was coupled directly to
the area element.

As I mentioned at the beginning, the {\em total} holonomy for a
surface is always trivial, or Abelian, for all these constructions
because otherwise surface ordering becomes ambiguous. The real
issue is to find a way of constructing a non-Abelian surface
holonomy in which the gauge field is not pure gauge. In the
construction where a two-form $B'$ couples to the area element and a
1-form $A$ couples to the edges, the total holonomy for
transporting a {\em string} along an infinitesimal rectangle is $F
+ B'\,,$ which has to vanish~\cite{Girelli:2003ev}\, if surface
ordering is to be unambiguous. In the construction with two
connections, the total holonomy for transporting a {\em field}
around the same rectangle is $F + \half {\rm d}_A V\,,$ which can
vanish even when the connection $A$ is not flat.

To sum up, I have constructed a gauge theory of Lie 2-groups, or
Lie crossed modules, by considering parallel transport of fields
around an infinitesimal rectangle with distinguished edges (string
configurations). Fields are transported along the string and away
from the string using different connections $A$ and $\bar A$, but
the result of transporting around a loop is still unambiguous
provided Eq.~(\ref{1202.weak}) is satisfied. The same condition is
required if parallel transport along two infinitesimally near
configurations of an infinitesimal string are gauge equivalent.
Summing over contributions from all such pairs of connections (or
gauge transformations) leads to an effective field theory of
non-Abelian two-forms.	In the absence of strings, there are no
distinguished edges, thus no integrability condition, and only
standard gauge theory remains. This suggests that in the presence
of strings or flux tubes of usual gauge theory, the effective
degrees of freedom are encoded in the gauge theory of a Lie crossed
module.

\appendix





\begin{thebibliography}{0}
\bibitem{Teitelboim:1986ya}
C.~Teitelboim,
Phys.\ Lett.\ B {\bf 167}, 63 (1986).

\bibitem{Breen:2001ie}
L.~Breen and W.~Messing,
``Differential Geometry of Gerbes,''
arXiv:math.ag/0106083.

\bibitem{Attal:2001}
R.~Attal,
``Two-dimensional parallel transport: combinatorics and
functoriality,''
arXiv:math-ph/0105050.

\bibitem{Attal:2002ix}
R.~Attal,
``Combinatorics of non-Abelian gerbes with connection and curvature,''
arXiv:math-ph/0203056.

\bibitem{Chepelev:2001mg}
I.~Chepelev,
JHEP {\bf 0202}, 013 (2002)

\bibitem{Baez:2002jn}
J.~C.~Baez,
``Higher Yang-Mills theory,''
arXiv:hep-th/0206130.

\bibitem{Hofman:2002ey}
C.~Hofman,
``Nonabelian 2-forms,''
arXiv:hep-th/0207017.

\bibitem{Pfeiffer:2003je}
H.~Pfeiffer,
Annals Phys.\  {\bf 308}, 447 (2003)

\bibitem{Girelli:2003ev}
F.~Girelli and H.~Pfeiffer,
J.\ Math.\ Phys.\  {\bf 45}, 3949 (2004).

\bibitem{Sharpe:2000ki}
E.~R.~Sharpe,
Phys.\ Rev.\ D {\bf 68}, 126003 (2003).


\bibitem{Vilenkin:1986ku}
A.~Vilenkin and T.~Vachaspati,
Phys.\ Rev.\ D {\bf 35}, 1138 (1987).


\bibitem{Lee:1993ty}
K.~Lee,
Phys.\ Rev.\ D {\bf 48}, 2493 (1993).




\end{thebibliography}
\end{document}